\documentclass[twocolumn,aps]{revtex4}
\usepackage{amssymb,graphicx}
\newcommand{\be}{\begin{equation}}
\newcommand{\ee}{\end{equation}}
\newcommand{\bea}{\begin{eqnarray}}
\newcommand{\eea}{\end{eqnarray}}
\newcommand{\ba}{\begin{array}}
\newcommand{\ea}{\end{array}}

\newcommand{\eq}[1]{(\ref{#1})}
\begin{document}
\title{Growth of epitaxial nanowires by controlled coarsening of strained islands}
\author{V.~B.~Shenoy}
\affiliation{Division of Engineering, Brown University,
Providence, RI 02912}
\date{\small \today}

\begin{abstract}
We show that elongated nanowires can be grown on crystal surfaces by allowing large strained two-dimensional islands to desorb by varying the adatom supersaturation or chemical potential. The width of the wires formed in this process is determined by a competition between the repulsive elastic interactions of the long edges of the wires and the thermodynamic driving force which tends to decrease the distance between these edges. The proposed mechanism allows for control of the wire sizes by changing the growth conditions, in particular, the vapor pressure of the material that is being deposited.
\end{abstract}
\maketitle

Strained heteroepitaxial growth of semiconductor and metal films has been of considerable interest in the recent years as it provides a versatile route to fabrication of nanostructures that have potential applications as electronic, optoelectronic and memory devices. Well studied examples of such structures include 3D islands or quantum dots in strained group IV and III-V systems. Mismatch strain can also give rise to elongated wire-like shapes with heights in the single digit nanometer range, since the increased perimeter of these structures leads to efficient relaxation of strain \cite{Tersoff,Brongersma,Batzill,Li,Zeilasek,Chen}. While nanowires may have application as non-lithographically fabricated interconnects, they can also be used to gain insight into fundamental physics of low-dimensional quantum systems. 

Although a competition between elastic and  boundary energies of 2D islands can give rise to elongated shapes,
it is difficult to control the island widths and lengths in an independent manner, since the aspect ratios are closely related to their area \cite{Li}. Recent experiments have also shown that strain in 2D islands can be also efficiently relaxed by ramification of the edges \cite{Muller} or by formation of curved boundaries \cite{Thayer}.  In this article, we show that elongated nanowires can be formed during the dissociation of large islands as a result of the competition between the repulsive elastic interactions between the edges of the island and thermodynamic driving force which tends to decrease the distance between the edges. Key advantages of the proposed  mechanism are that 1) it allows the control of the width of the wire by changing the chemical potential or vapor pressure of the adatom reservoir and 2) elongated nanowires can also be formed in systems where ramified or curved boundaries are responsible for stress relaxation of large islands \cite{Muller,Thayer}.

\begin{figure}[h!]
\begin{center}
\includegraphics[width=0.18\textwidth,height=0.24\textwidth ]{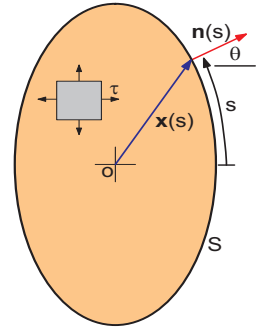}
\end{center}
\vspace{-0.5cm}
\caption{ Schematic of the strained island or surface domain, whose boundary is represented by $S$, the Cartesian coordinate and the unit-normal of a point on the boundary is denoted by ${\bf x}(s)$ and ${\bf n}(s)$, respectively, where $s$ is the arc-length. In the case of a strained island, $\tau$ represents the product of the equibiaxial mismatch stress and the height of the island.}    
\label{fig1}
\vspace{-0.5cm}
\end{figure}

If island growth is limited by the attachment of atoms to its perimeter, the outward normal velocity can be written as
\be
v_n(s) = k\left[\Delta \mu - \kappa({\beta}(\theta)+\beta^{\prime \prime}(\theta)) - \tau \nabla \cdot {\bf u}(s)\right],
\ee
where $s$ denotes the arc-length (refer to Fig.~1), $k$ is related to the mobility of the perimeter, $\Delta \mu$ gives the driving force on a straight edge, $\kappa$ is the curvature (positive if the center of curvature is inside the island),  $\beta(\theta)$ is the energy required to create a unit length of the island boundary, and the last term is the contribution due to elastic relaxation of material close to the island perimeter. Here, $\tau$ is the product of the equibiaxial mismatch stress \cite{Anisotropy} and the height of the strained island, or the difference between the surface stresses of the domain and the substrate in the case of strained surface domains \cite{Thayer}. The divergence of the displacement field can be written using elastic Green's functions \cite{Landau} as
\be
\nabla \cdot{\bf u}(s) = \frac{(1-\nu^2)\tau}{\pi E}\int_{S} \frac{\left[{\bf{x}}(s)-{\bf{x}}({s}^{\prime})\right]\cdot{\bf{n}}({s}^{\prime})}{\left|{\bf{x}}(s)-{\bf{x}}({s}^{\prime})\right|^3}ds^{\prime},
\ee
where $\nu$ and $E$ are the Poisson's ratio and the Young's modulus of the substrate, respectively, and the Cartesian coordinate ${\bf x}$ and the outward unit-normal ${\bf n}$ are shown in Fig.~1. In what follows, we  study shape evolution of islands by discretization of their perimeter into small linear segments and by moving these segments according to the velocity in Eq.~(1). The integrand in Eq.~(2), however, is singular due to the divergent nature of the elastic fields at the island perimeter. It can be regularized using an elastic cut-off length $\delta$, which we have set to 2 \AA~in the numerical calculations.

\begin{figure}[h!]
\begin{center}
\includegraphics[width=8.2cm ]{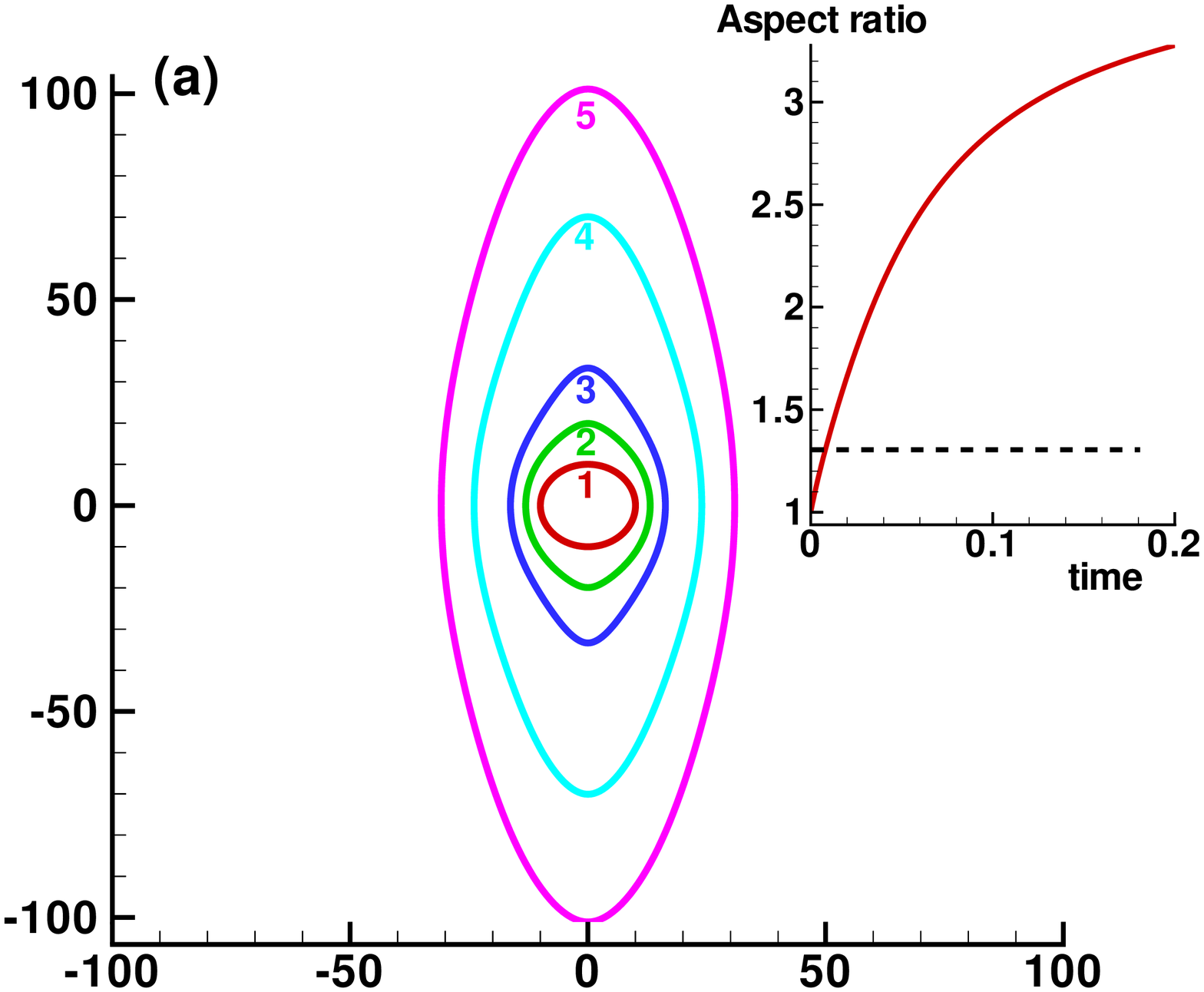}
\end{center}
\vspace{-1.8cm}
\begin{center}
\includegraphics[width=8.2cm]{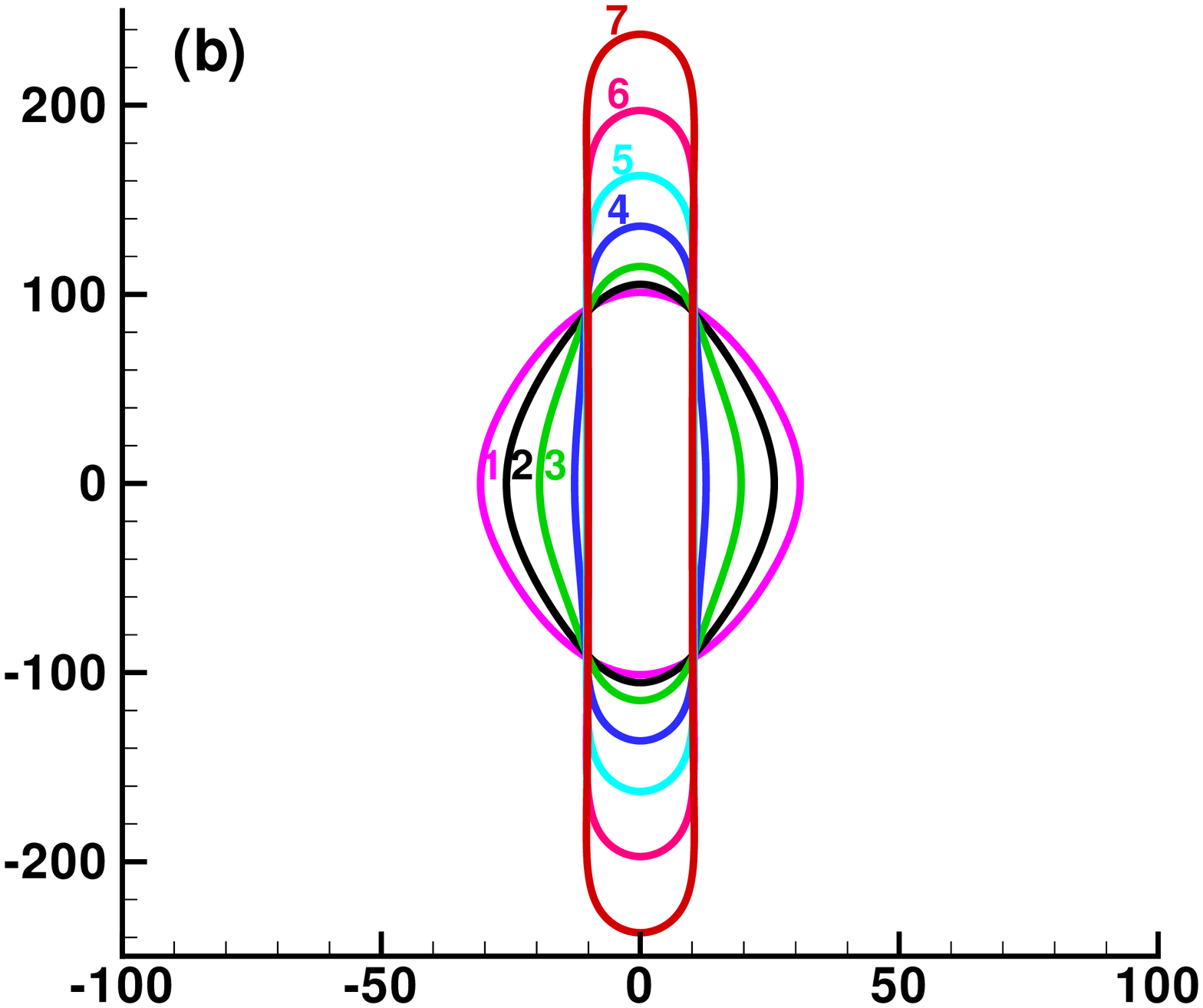}
\end{center}
\vspace{-1cm}
\caption{(a) Growth shapes of a strained island with a 2-fold symmetric boundary energy. The red curve in the inset graph shows the evolution of the island aspect ratio (ratio of the lengths of major and the minor axes), while the black dashed curve denotes the equilibrium aspect ratio when strain effects are not included. (b) Dissociation shapes of the island marked 5 in (a). When $\Delta \mu <0$, the island initially shrinks in the horizontal direction before it begins to steadily grow along the vertical direction. All the lengths are given in $nm$. }    
\label{fig2}
\vspace{-0.5cm}
\end{figure}

We first consider the growth of islands in a system with a 2-fold symmetric boundary energy, 
$\beta(\theta) = \beta_0(1-\epsilon \cos 2 \theta)$, with the anisotropy factor $\epsilon$ = 0.1. The effect of strain on the island shapes can be conveniently discussed in terms of the dimensionless parameter $\alpha = F/\beta_0$, where $F = (1-\nu^2)\tau^2/(\pi E)$. 
In the case of surface domains studied in \cite{Thayer} and \cite{Hannon}, $\alpha \approx 0.3$, while it is smaller ($\alpha \approx 0.05$) in the case of 2-fold symmetric Si(001) islands considered in \cite{Zeilasek}. The growth shapes ($\Delta \mu > 0$) of a circular nucleus for $\alpha = 0.2$ are shown in Fig.~2(a). We find that the aspect ratio  continously increases as the island grows in size, far exceeding the aspect ratio of the equilibrium shape without strain effects (refer to the inset in Fig.~1(a)). The evolution of the aspect ratios is qualitatively similar to the experimental observations of island growth on Si(001) by Zeilasek et al \cite{Zeilasek}. We can understand the fast outward growth of the regions with smaller boundary stiffness by comparing the contributions to the normal velocity from elastic and line-tension effects. 
While the contribution from the line energy is proportional to the curvature $\kappa$, the elastic contribution is opposite in sign and can be shown to be proportional to $\kappa \log(\kappa \delta)$. Therefore, regions with large curvature and small boundary stiffness tend to move outwards at a more rapid rate compared to the regions of smaller curvature due to larger contribution from the elastic relaxation effects in the former case.

 Next, we consider the evolution of the shape marked 5 in Fig.~2(a), when the driving force $\Delta \mu < 0$. In this case, as the material close to the minor (shorter) axis of the island begins to desorb, the width of the island decreases, while the regions near the major axis show a very small outward movement (refer to shapes 1-3 in Fig.~2(b)). While the contributions of the line energy and elastic relaxation to the normal velocity are smaller than the contribution from the driving force near the minor axis, its effect is offset to a large extent by the elastic contribution near the major axis due to the higher curvature of this region.
As the island shape evolves, the width of the island steadily decreases and eventually reaches a point where the repulsive elastic interaction between its nearly straight edges (shape 4 in Fig.~2(b)) becomes comparable to the driving force, $\Delta \mu$. At this point, the competition between these two effects leads to an optimum width of the island, while the top and bottom ends continue steadily move in an outward direction giving the elongated shapes shown in Fig.~2(b). An animation of the wire formation process can be found at the web-site given in Ref.~\cite{Web}.

A quantitative expression for the width $w$ of the wire can be derived by considering the equilibrium configuration of an infinitely long wire for $\Delta \mu < 0$. Since the velocity of the straight edges of the wire
must vanish in equilibrium, we have,
\be
v_n = k\left(\Delta \mu + \frac{2F}{w}\right) = 0, \,\,\, {\mathrm {or}} \,\,\, w = \frac{2F}{-\Delta \mu}.
\label{width}
\ee
In the calculations in Fig.~2(b), we have chosen $\Delta \mu = -0.004 \beta_0/\delta$ and $F = 0.2 \beta_0$, which gives $w = 100 \delta$, in agreement with the observed width of the wire. Since the width of the wire depends on $\Delta \mu$, it can be controlled by varying the supersaturation or the vapor pressure of the material that is being grown. The competition that leads to the wire-like shapes in Fig.~2(b) also occurs in the case of elongated stripes of 7x7 domains of Si(111) that coexist with the 1x1 domains \cite{Hannon} at temperatures where the free energy of the former structure is larger than the latter (which is equivalent to $\Delta \mu < 0$). As in Eq.~\eq{width}, the stripes are stabilized by the repulsive elastic interactions between the edges. It should also be noted that since Eq.~\eq{width} does not depend on the form of boundary energy, elongated shapes could also be achieved in systems with other types of symmetry, as illustrated in the next paragraph. 
\begin{figure}[h!]
\vspace{-0.5cm}
\begin{center}
\includegraphics[width= 8.2cm ]{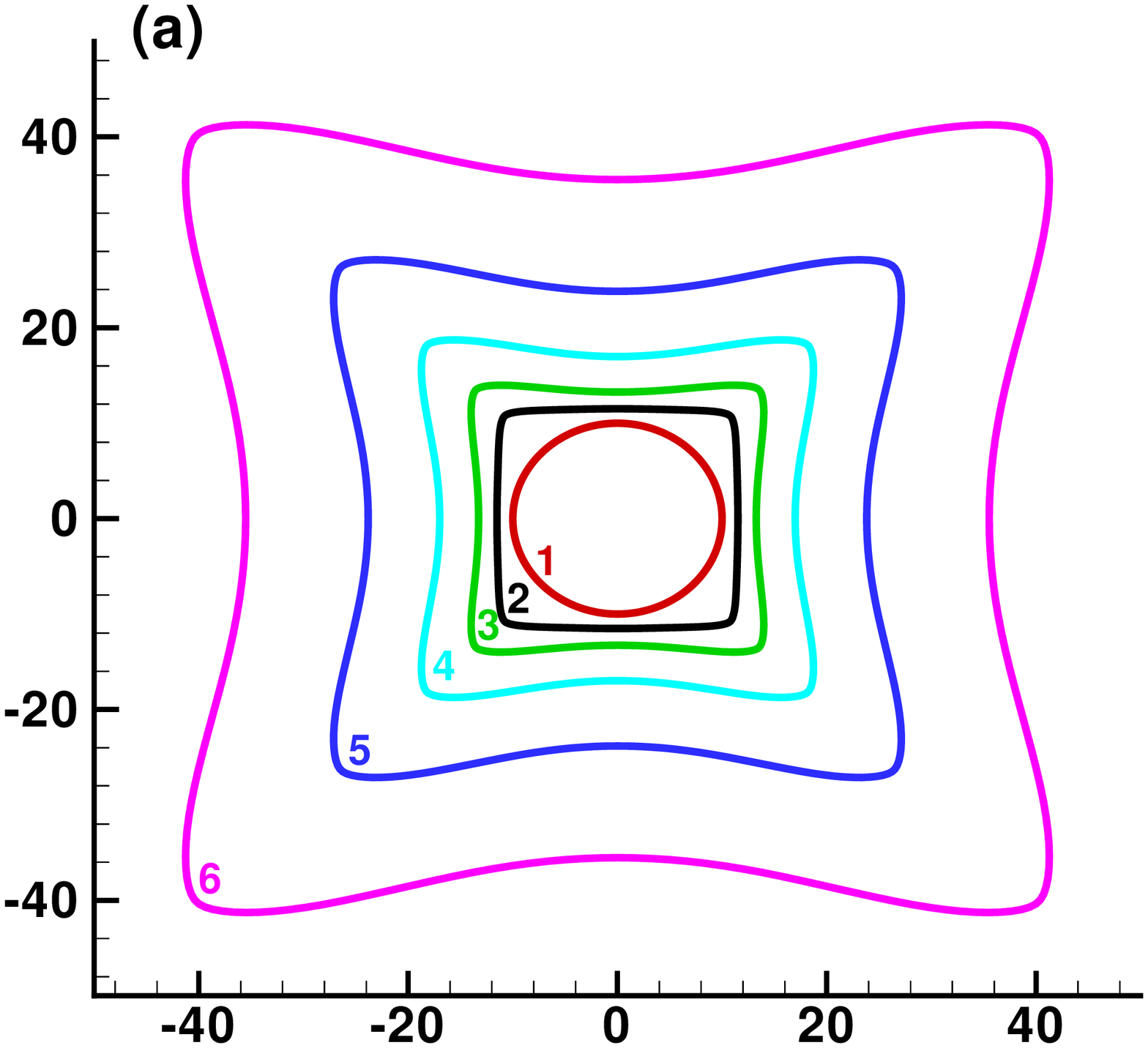}
\end{center}
\vspace{-1.8cm}
\begin{center}
\includegraphics[width= 8.2 cm]{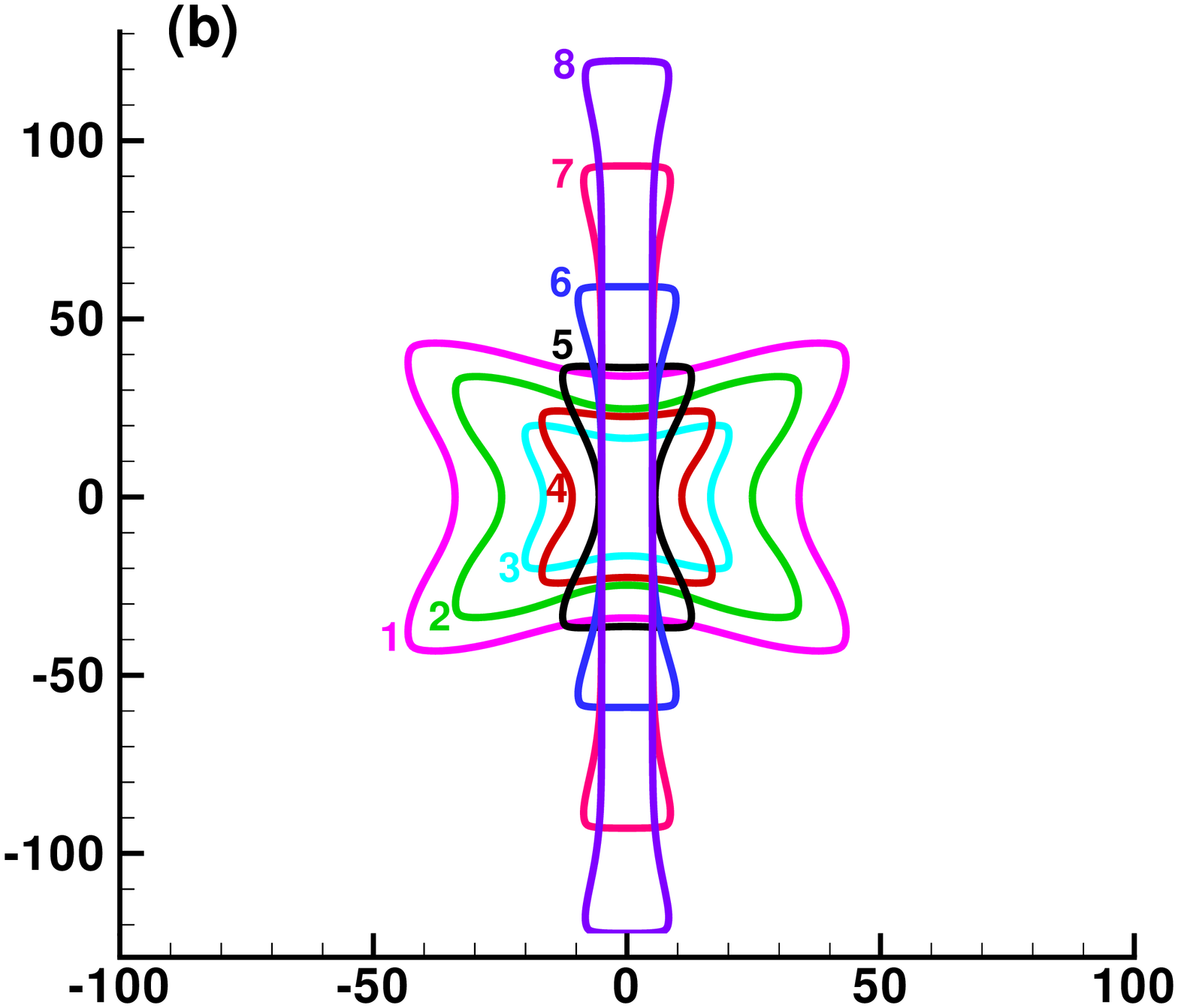}
\end{center}
\vspace{-1cm}
\caption{(a) Growth shapes of a strained island with a 4-fold symmetric boundary energy. Note that as the island grows in size, its edges become more concave in curvature. (b) Dissociation shapes of the island marked 6 in (a). When $\Delta \mu <0$, the island initially shrinks in size before it begins to steadily grow along the vertical direction. All the lengths are given in $nm$. }    
\label{fig3}
\vspace{-0.5cm}
\end{figure}

In Fig.~3(a), we show the growth shapes of an island with a 4-fold symmetric boundary energy, $\beta(\theta) = \beta_0(1- 0.033 \cos 4 \theta$), for $F/\beta_0 = 0.2$. Here, as the circular nucleus grows in size, its shape initially evolves to a square, after which the boundaries acquire concave features with curvatures that continuously increase with size. Such concave shapes, which lead to relaxation of mismatch strain, have been very recently observed on strained domains on Si(111) \cite{Thayer}. The dissociation shapes  of the island marked 6 in Fig.~3(a), with $\Delta \mu = - 0.008 \beta_0/\delta$, are shown in Fig.~3(b). The shape evolution involves two stages: First, the island shrinks in size until it reaches the shape marked 3 in Fig. 3(b) at which point the system slowly elongates in the vertical direction (refer to the animation in Ref.~\cite{Web}) due to the instability discussed in Refs.~\cite{Tersoff} and \cite{Li}. As the island elongates at fixed $\Delta \mu$, the curvatures close to top and bottom edges become large, leading a steady outward propagation, while the long edges become straighter and achieve the optimum width given by Eq.~\eq{width}. It should be pointed out that the shape marked 3 is a shallow metastable state and the direction of elongation (horizontal or vertical) depends on the small noise in the simulations due to finite arithmetic precision. In an experimental situation, thermal noise will select the direction of propagation.

While most experiments to date focus on growth or equilibrium shapes of islands, the shapes of strained Ca/CaF$_2$ islands observed by Batzill and Snowdon \cite{Batzill} can perhaps be understood on the basis of the present work. While island shapes on atomically smooth CaF$_2$ surfaces are consistent with the equilibrium theory in Ref.\cite{Tersoff}, the widths of the long wire-like shapes on stepped surfaces were found to be significantly larger than the predictions of this model. A possible explanation for the discrepancy is that the step-edges can provide favorable sites for attachment of Ca atoms, and therefore give rise to an effective $\Delta \mu < 0$, which would lead to coarsening of large islands. As the large islands lose Ca atoms, the competition between elastic interactions and the coarsening effect can result in the observed wire-like shapes. Of course, more experimental and theoretical work is required for a quantitative verification of this argument. 

In conclusion, we have developed a numerical method which gives growth shapes of strained domains that are similar to the shapes observed in recent experiments. Using this method we have shown that straight and elongated nanowires can be grown by controlled desorption of large islands. 

The research support of the NSF
through grants CMS-0093714 and CMS-0210095 and the Brown University
MRSEC program (DMR-0079964) is gratefully acknowledged.


\begin{thebibliography}{30}

\bibitem{Tersoff}
J.~Tersoff and R.~M.~Tromp, Phys. Rev. Lett. {\bf 70} 2782 (1993).

\bibitem{Brongersma}
S.~H.~Brongersma, M.~R.~Castell, D.~D.~Perovic and M.~Zinke-Allmang, Phys. Rev. Lett. {\bf 80}, 3795 (1998).

\bibitem{Batzill}
M.~Batzill and K.~J.~Snowdon, Appl. Phys. Lett. {\bf 77}, 1955 (2000).

\bibitem{Li}
A.~Li, F.~Liu and M.~G.~Lagally, Phys. Rev. Lett. {\bf 85} 1922 (2000).

\bibitem{Zeilasek}
V.~Zeilasek, F.~Liu, Y.~Zhao, J.~B.~Maxson and M.~G.~Lagally, Phys. Rev. B {\bf 64}, 201320 (2001).

\bibitem{Chen}
Y.~Chen, D.~A.~A.~Ohlbererg and R.~S.~Williams, J. Appl. Phys. {\bf 91} 3213 (2002).

\bibitem{Muller}
B.~M\H{u}ller, L.~Nedelmann, B.~Fischer, H.~Brune, J.~V.~Barth and K.~Kern, Phys. Rev. Lett. {\bf 80} 2642 (1998).

\bibitem{Thayer}
G.~E.~Thayer, J.~B.~Hannon and R.~M.~Tromp, Nature Materials {\bf 3}, 75 (2004).

\bibitem{Anisotropy}
The growth of islands with aniosotropic mismatch strains (e.g. silicide wires in Ref.\cite{Chen}) will be considered in a forthcoming article.

\bibitem{Landau}
L.~D.~Landau and E.~M.~Lifshitz, {\em Theory of Elasticity}, Third Edition, (Pergamon,1986) p. 25.

\bibitem{Hannon}
J.~B.~Hannon, F.-J.~Meyer zu Heringdorf, J.~Tersoff and R.~M.~Tromp, Phys. Rev. Lett. {\bf 86}, 4871 (2001).

\bibitem{Web} 
http://www.cascv.brown.edu/$\sim$shenoy/nanowire.html


\end{thebibliography}
\end{document}